\documentclass[
 a4paper,
]{jpconf}
\usepackage{graphicx}
\usepackage{amsmath}
\usepackage{amssymb}
\usepackage{bm}

\begin{document}
\title{Electrical conductivity in helical and conical magnetic states}

\author{Shun~Okumura$^1$, Takahiro~Morimoto$^2$, Yasuyuki~Kato$^2$, and Yukitoshi~Motome$^2$}

\address{$^1$The Institute for Solid State Physics, The University of Tokyo, Kashiwa 277-8581, Japan}
\address{$^2$Department of Applied Physics, The University of Tokyo, Tokyo 113-8656, Japan}

\ead{okumura@issp.u-tokyo.ac.jp}

\begin{abstract}
We theoretically study the electrical conductivity in a one-dimensional helimagnet whose spin texture changes from helimagnetic to conical magnetic, and to forced ferromagnetic state while increasing the magnetic field along the helical axis.
We find that the conductivity in the helimagnetic state at zero field depends on the electron filling and the coefficient of the spin-charge coupling.
We also find that the conductivity in the conical magnetic state changes nonlinearly to the applied field, and the magnetoresistance becomes negative and positive depending on the model parameters.
\end{abstract}

\section{\label{s1}Introduction}

\begin{figure}[t]
\centering
\includegraphics[width=\linewidth,clip]{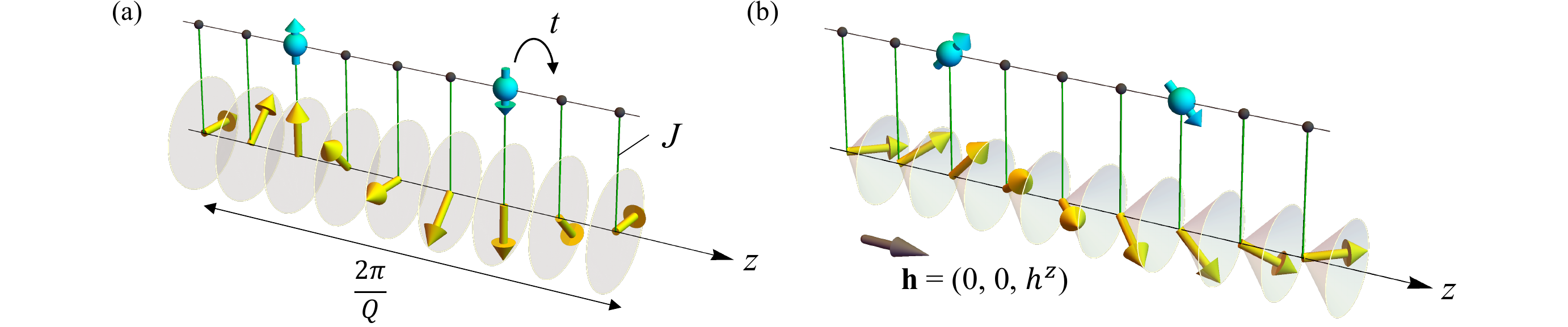}
\caption{
Schematic pictures of the model~\eqref{eq:H} in the (a) helical and (b) conical magnetic states with the helical pitch $\frac{2\pi}{Q}$. 
The localized spins (yellow) are coupled to spins of itinerant electrons (cyan) by the coupling constant $J$.
The gray arrow in (b) represents the magnetic field parallel to the helical axis, which is taken as the $z$ axis.
}
\label{f1}
\end{figure}

A helical magnetic state (HM) is an archetype of the noncollinear magnetic texture, which has a periodically twisted spin structure, as shown in figure~\ref{f1}(a).
The HM has attracted considerable attention in condensed matter physics since they give rise to peculiar physical phenomena in multiferroics~\cite{Tokura2014} and spintronics~\cite{Yang2021} due to lack of the spatial-inversion symmetry.
In an external magnetic field parallel to the helical axis, the HM turns into a noncoplanar magnetic texture, which is called the conical magnetic state (CM) [figure~\ref{f1}(b)], and finally relaxes to a forced ferromagnetic state (FFM).
Recently, the CM in the magnetic conductors has attracted many interests due to the peculiar electrical transport, for instance, the magnetoresistance effect~\cite{Yonemura_etal2017} and the electric magnetochiral effect~\cite{Aoki2019, Jiang2020}.
While the spin transport was studied in both HM and CM~\cite{Watanabe2016, Okumura2019, Ustinov2020PMM}, the charge transport has not been fully clarified yet.

In this paper, we report our theoretical study on the electronic transport properties in a one-dimensional helimagnet.
We calculate the electrical conductivity by the linear response theory varying the spin-charge coupling constant, the electron filling, and the magnetization.
We find that the conductivity in the HM at zero field is maximized around quarter filling similar to the FFM case, whereas the peak shifts to higher filling as the spin-charge coupling decreases.
We also find that the CM in the magnetic field exhibits not only negative but also positive magnetoresistance depending on the model parameters.

\section{\label{s2}Model}

We consider a one-dimensional spin-charge coupled model, whose Hamiltonian is given by 
\begin{equation}
	H = -t\sum_{l, \mu}(c^{\dagger}_{l\mu}c^{\;}_{l+1\mu}+\mathrm{h.c.})-J\sum_{l, \mu,\nu}c^{\dagger}_{l\mu}{\boldsymbol \sigma}_{\mu\nu}c^{\;}_{l\nu}\cdot{\mathbf S}_l -h^z\sum_lS_l^z,
	\label{eq:H}
\end{equation}
where $c_{l\mu}$ ($c^{\dagger}_{l\mu}$) is an annihilation (creation) operator for an itinerant electron with spin $\mu=\uparrow,\downarrow$ at site $l$, and $\boldsymbol{\sigma} = (\sigma^x,\sigma^y,\sigma^z)$ is the vector of Pauli matrices.
The first term describes the kinetic energy of itinerant electrons with the nearest-neighbor hopping $t$ and the second term represents the onsite coupling between the itinerant electrons and the localized classical spins $\bold{S}_l$ with the coupling constant $J$.
The last term is the Zeeman coupling to the external magnetic field along the chain direction (the $z$ axis) which is taken into account only for the localized spins for simplicity. 
We show schematic pictures of the model~\eqref{eq:H} in figure~\ref{f1}.

We study the ground state of the model~\eqref{eq:H} by assuming a swirling magnetic texture for the localized spins, ${\mathbf S}_{l}=(\sqrt{1-m^2}\cos(Ql), \sqrt{1-m^2}\sin(Ql), m)$, which describes the HM at $m=0$ [figure~\ref{f1}(a)], the CM for $0<m<1$ [figure~\ref{f1}(b)], and the FFM at $m=1$;
$m$ represents the magnetization per spin induced by the external magnetic field, and $Q$ is the wavenumber specifying the helical pitch, $2\pi/Q$.
By the Fourier transformation of \eqref{eq:H}, we obtain two energy bands split by $\sim2J$~\cite{Okumura2021}, whose energy dispersions are given by
\begin{equation}
	\varepsilon_{\lambda}({k}) = -2t\cos\frac{Q}{2}\cos{k}+\lambda\sqrt{4t^2\sin^2\frac{Q}{2}\sin^2{k}+4tJm\sin\frac{Q}{2}\sin{k}+J^2}, 
\label{eq:dispersion}
\end{equation}
where $\lambda=+$ and $-$ represent the higher- and lower-energy bands, respectively, and $k$ is the shifted wavenumber defined as $k = \tilde{k}+\frac{Q}{2}\lambda$, where $\tilde{k}$ is the original wavenumber. 
Here and hereafter, we take the lattice constant unity.

In the strong coupling limit $J\rightarrow\infty$, the spins of itinerant electrons are parallel to the localized spins at each site, and hence, the effective hopping amplitude depends on the relative angle between the neighboring localized spins~\cite{Anderson1955}.
For the swirling spin configuration, the energy dispersions are asymptotically described by the simple cosine form as 
\begin{equation}
	\varepsilon_{\lambda}({k}) = -2t\sqrt{1-(1-m^2)\sin^2\frac{Q}{2}}\cos(k-\varphi)+\lambda J,
\label{eq:double_exchange}
\end{equation}
where $\varphi$ satisfies $\tan\varphi=m\lambda\tan\frac{Q}{2}$.

\section{\label{s3}Linear response theory}

We calculate the electrical conductivity for the model~\eqref{eq:H} by using the linear response theory as
\begin{equation}
	\sigma = \frac{e^2}{\hbar}\int\frac{dk}{2\pi}\left[\tau\sum_\lambda\left(\frac{\partial f_\lambda}{\partial\varepsilon_\lambda}\right)\langle \lambda|\partial_{k}H|\lambda\rangle^2+i\sum_{\lambda\neq\lambda'}\frac{f_\lambda-f_{\lambda'}}{\varepsilon_{\lambda'}-\varepsilon_\lambda}\frac{\langle\lambda|\partial_{k}H|\lambda'\rangle\langle\lambda'|\partial_{k}H|\lambda\rangle}{\varepsilon_\lambda-\varepsilon_{\lambda'}+i/\tau}\right],
	\label{eq:linear_response}
\end{equation} 
where $|\lambda\rangle$ and $f_\lambda=f(\varepsilon_{\lambda})$ are the eigenvector and the Fermi distribution function, respectively; $\tau$ represents the relaxation time, which is assumed to be a positive constant for simplicity.
The first and second terms in \eqref{eq:linear_response} represent the Drude and interband contributions, respectively.
In the ground state, we can ignore the second term for $J\geq2t$ because the two energy bands in~\eqref{eq:dispersion} do not overlap with each other. 
Then, \eqref{eq:linear_response} is reduced to
\begin{equation}
	\sigma = \frac{e^2\tau}{\pi\hbar}\sum_{\lambda,j}\left|t\cos\frac{Q}{2}\sin k^\lambda_{\mathrm{F}j}+\lambda\frac{t^2\sin^2\frac{Q}{2}\sin(2k^\lambda_{\mathrm{F}j})+tJm\sin\frac{Q}{2}\cos{k^\lambda_{\mathrm{F}j}}}{\sqrt{4t^2\sin^2k^\lambda_{\mathrm{F}j}\sin^2\frac{Q}{2}+4tJm\sin\frac{Q}{2}\sin{k^\lambda_{\mathrm{F}j}}+J^2}}\right|,
	\label{eq:Drude}
\end{equation} 
where $k^\lambda_{\mathrm{F}j}$ represents the $j$th Fermi wavenumber of $\varepsilon_\lambda(k)$.
Furthermore, in the limit of $J\rightarrow\infty$, using \eqref{eq:double_exchange}, we obtain
\begin{equation}
	\sigma = \sqrt{1-(1-m^2)\sin^2\frac{Q}{2}}\sigma_\mathrm{FFM},
\label{eq:J_inf}
\end{equation}
where $\sigma_\mathrm{FFM}=\frac{e^2\tau}{\pi\hbar}2t|\sin(\pi n)|$ represents the electrical conductivity in the FFM at $m=1$;
$n$ is the electron filling (the average number of electrons per site).
In the following calculation, we focus on the situation when the Fermi energy is in the lower-energy band and $J\geq2t$, i.e., $0<n<1$, and we set $t=1$ as the energy unit.

\section{\label{s4}Calculation results}

\begin{figure}[b]
\centering
\includegraphics[width=\columnwidth,clip]{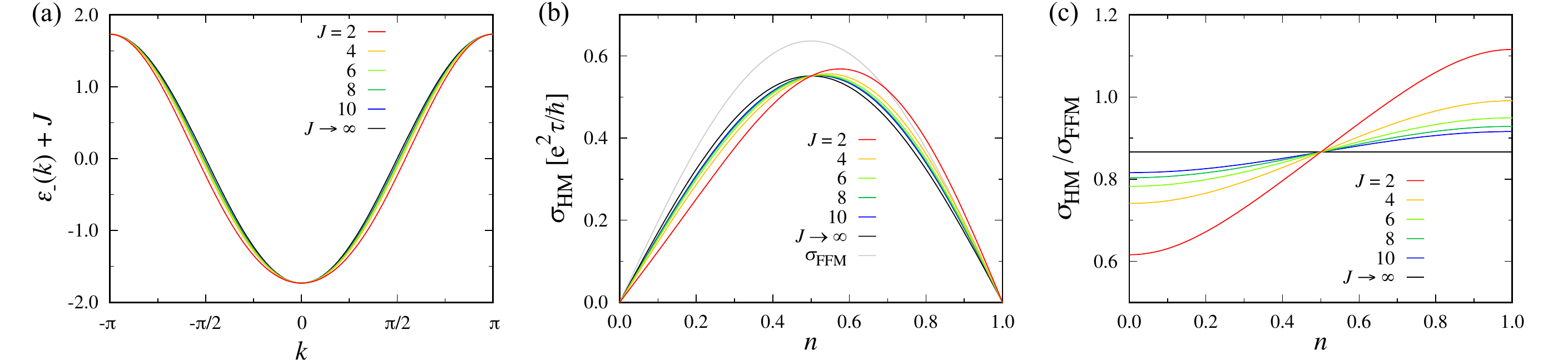}
\caption{
(a) Energy dispersion for the lower-energy band, $\varepsilon_{-}(k)$, plotted with the offset of $J$, and (b) $n$ dependence of the electrical conductivity in the helical magnetic state at $m=0$, $\sigma_\mathrm{HM}$, for several $J$.
We take $Q=\frac{2\pi}{6}$. 
The gray curve shows the data in the forced ferromagnetic state at $m=1$, $\sigma_\mathrm{FFM}$.
(c) $n$ dependence of $\sigma_\mathrm{HM}/\sigma_\mathrm{FFM}$.
}
\label{f2}
\end{figure}

First, we show the results for the HM at $m=0$ in figure~\ref{f2}.
As shown in figure~\ref{f2}(a), the energy dispersion $\varepsilon_{-}(k)$ deviates from the cosine curve in \eqref{eq:double_exchange} as $J$ decreases.
Since $\varepsilon_{-}(k)$ is symmetric with respect to $k$ at $m=0$, the Fermi wavenumber is given by $n$ as $k^{-}_{\mathrm{F}1}=-k^{-}_{\mathrm{F}2}=\pi n$ for $\frac{2\pi}{Q}\geq4$.
We note that $\varepsilon_-(k)$ shows a double-well structure when the helical pitch becomes shorter and satisfies $2\pi/Q <4$, while it is not shown in figure~\ref{f2}(a).

In figure~\ref{f2}(b), we plot the electrical conductivity in the HM, $\sigma_\mathrm{HM}$, calculated by \eqref{eq:Drude} at $Q=\frac{2\pi}{6}$ as functions of $n$ for several values of $J$.
At $m=0$, the first term in \eqref{eq:Drude} corresponds to the strong coupling limit in \eqref{eq:J_inf}, and the second term represents the deviation from the strong coupling limit.
We find that $\sigma_\mathrm{HM}$ is a symmetric dome-shaped function of $n$ in the limit of $J\to\infty$ similar to $\sigma_\mathrm{FFM}$, while the peak is shifted from $n=0.5$ to higher $n$ as $J$ decreases due to the modulation of $\varepsilon_{-}(k)$ in figure~\ref{f2}(a).
Note that the second term in \eqref{eq:Drude} vanishes irrespective of $J$ at $n=0.5$, and hence the electrical conductivity is constant as $\sigma_\mathrm{HM}=\frac{2te^2\tau}{\pi\hbar}\cos\frac{Q}{2}$.

We also plot the ratio $\sigma_\mathrm{HM}/\sigma_\mathrm{FFM}$ in figure~\ref{f2}(c).
In the limit of $J\to\infty$, $\sigma_\mathrm{HM}/\sigma_\mathrm{FFM}=\cos\frac{Q}{2}$ irrespective of $n$, while $\sigma_\mathrm{HM}/\sigma_\mathrm{FFM}$ becomes smaller (larger) as decreasing $J$ for $n<0.5$ ($n>0.5$). 
Interestingly, the ratio can be larger than $1$ in the high $n$ and small $J$ region, which indicates unusual positive magnetoresistance. 

\begin{figure}[h]
\centering
\includegraphics[width=\columnwidth,clip]{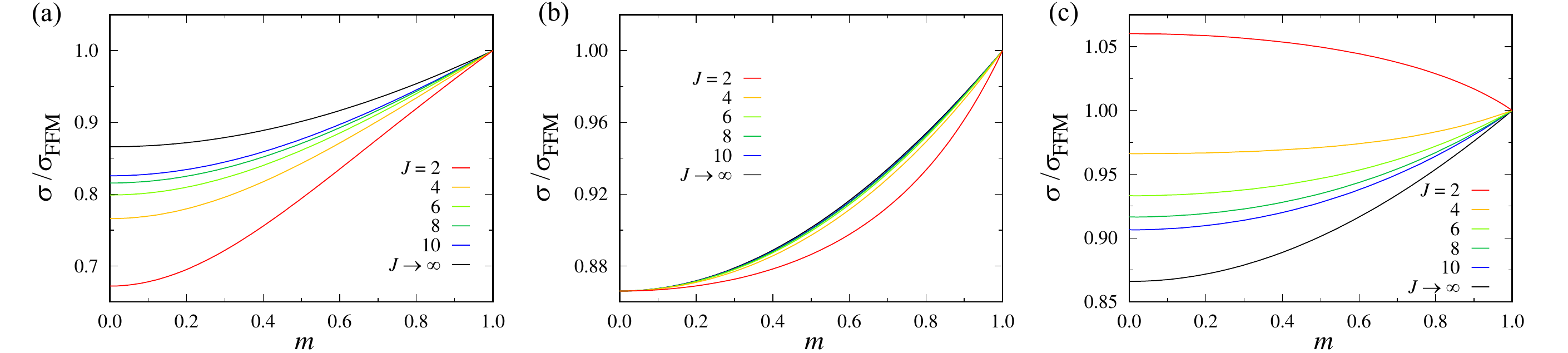}
\caption{
$m$ dependence of the electrical conductivity normalized by the value in the FFM, $\sigma/\sigma_\mathrm{FFM}$, at (a) $n=0.2$, (b) 0.5, and (c) 0.8 for $Q=\frac{2\pi}{6}$ and several $J$.
}
\label{f3}
\end{figure}

In figure~\ref{f3}, we show the magnetization dependence of the electrical conductivity, whose inverse gives the magnetoresistance.
The energy dispersion is asymmetrically distorted in the CM for $0<m<1$~\cite{Okumura2021}, and the Fermi wavenumber is no longer symmetric with respect to $k$, $k^{-}_{\mathrm{F}1}\neq-k^{-}_{\mathrm{F}2}$. 
This distortion gives rise to the further modulation of $\sigma$ as given by the second term of \eqref{eq:Drude}.
Figures~\ref{f3}(a) and \ref{f3}(b) show the data for $n=0.2$ and $n=0.5$, respectively, indicating nonlinear negative magnetoresistance.
On the other hand, the results for $n=0.8$ in figure~\ref{f3}(c) show that the magnetoresistance changes from negative to positive while decreasing $J$, as suggested in figure~\ref{f2}(c).

\section{\label{s5}Summary}

In summary, we have studied the electrical conductivity in the HM and CM by using the linear response theory for the one-dimensional spin-charge coupled model.
We found that the conductivity in the HM at zero field becomes the dome-like shape as a function of the electron filling, while the functional form depends on the spin-charge coupling reflecting the modulation of the energy dispersion by the spiral magnetic texture.
We also found that the conductivity in the CM exhibits the nonlinear magnetic field dependence, including unusual positive magnetoresistance in the hole doped region from half filling.
Our results would be experimentally relevant to the quasi-one-dimensional helimagnets, e.g., CrNb$_3$S$_6$~\cite{Yonemura_etal2017}. 
Our findings give a starting point for further study of peculiar transport phenomena in magnets showing noncollinear and noncoplanar magnetic textures.

\ack{
This research was supported by JST CREST (Nos.~JPMJCR18T2 and JPMJCR19T3), JST PRESTO (No.~JPMJPR19L9), the JSPS KAKENHI (Nos.~JP19H05822 and JP19H05825).
}

\section*{References}
\providecommand{\newblock}{}

\end{document}